\documentclass[aps,prl,twocolumn,natbib,showpacs,floatfix,superscriptaddress]{revtex4-2}

\usepackage{graphicx}
\usepackage{amsmath,amssymb,bbold,bm,color}
\usepackage{float}
\usepackage{epstopdf}
\usepackage{hyperref}

\usepackage{color}
\usepackage{enumerate}
\usepackage{wasysym}
\usepackage{url}
\usepackage{dsfont}
\usepackage{braket}
\usepackage{comment} 
\usepackage{xcolor}

\newcommand{\bd}{\mathbf{d}}
\newcommand{\bdh}{\hat{\mathbf{d}}}

\newcommand{\bk}{\mathbf{k}}

\newcommand{\bA}{\mathbf{A}}

\newcommand{\bOmega}{\mathbf{\Omega}}

\newcommand{\bv}{\mathbf{v}}
\newcommand{\balpha}{\boldsymbol{\alpha}}
\newcommand{\bsigma}{\boldsymbol{\sigma}}

\newcommand{\kv}{\mathbf{k}}

\newcommand{\cP}{{\cal P}}

\newcommand{\cI}{{\cal I}}

\definecolor{IEcolor}{RGB}{255, 153, 51}

\renewcommand{\vec}[1]{{\bf{#1}}}

\hypersetup{colorlinks=true,linkcolor=blue,citecolor=blue,urlcolor=blue}

\begin{document}

\title{Topological frequency conversion in rhombohedral multilayer graphene}

\author{\'Etienne Lantagne-Hurtubise}
\affiliation{Department of Physics and Institute for Quantum Information and Matter, California Institute of Technology, Pasadena, California 91125, USA}

\author{Iliya Esin}
\affiliation{Department of Physics and Institute for Quantum Information and Matter, California Institute of Technology, Pasadena, California 91125, USA}

\author{Gil Refael}
\affiliation{Department of Physics and Institute for Quantum Information and Matter, California Institute of Technology, Pasadena, California 91125, USA}

\author{Frederik Nathan}
\affiliation{Department of Physics and Institute for Quantum Information and Matter, California Institute of Technology, Pasadena, California 91125, USA}
\affiliation{Center for Quantum Devices, Niels Bohr Institute, University of Copenhagen, 2100 Copenhagen, Denmark}

\date{\today}

\begin{abstract}
We show that rhombohedral multilayer graphene supports  
topological frequency conversion, whereby 
a fraction of electrons transfer energy between two monochromatic light sources at a quantized rate. The pristine nature and 
gate tunability of these materials, along with a
Berry curvature that directly couples to electric fields, make them ideal platforms for the
experimental realization of topological frequency conversion. Among the rhombohedral family, we find that Bernal bilayer graphene appears most promising for THz-scale applications due to lower dissipation. We discuss strategies to circumvent cancellations between the two valleys of graphene and to minimize dissipative losses using commensurate frequencies, thus opening a potential pathway for net amplification.
 \end{abstract}

\maketitle

Quantum systems coupled to light
exhibit a range of  unique  emergent
phenomena, providing a powerful platform to design and control 
phases of matter~\cite{Oka_2009,Kitagawa2010,Lindner_2011,Nathan_2015,Dehghani_2015,Titum_2016,Khemani_2016,Else_2016,Potter_2016,Roy_2017,Oka_2019,Rudner_2020,Nathan_2019b,Esin_2020, Yang_2023}. It was recently appreciated that the paradigm of Floquet engineering can be generalized
to  {\it multi-chromatic} driving, leading to even richer physics~\cite{Verdeny2016,Martin2017,Kolodrubetz_2018,Peng2018,Nathan2019, Crowley2019, Else_2020, Boyers2020,  Chen2020, Long_2020, Crowley2020, Nathan_2020c, Korber2020, Nathan2021a, Maltz_2021, Nathan2022, Schwennicke2022, Nathan2022}. In this setting, the photon numbers of distinct driving modes act as synthetic lattice coordinates, opening  access to higher-dimensional topological phases that support novel pumping mechanisms. A prime example is \emph{topological frequency conversion},
whereby energy is transferred between two modes with frequencies $f_1$ and $f_2$ at a rate given by the power quantum
$h f_1 f_2$. 

In the solid-state realm, Weyl semimetals were recently proposed as a  platform 
to realize this effect, leveraging the sources and sinks of Berry curvature carried by their Weyl points~\cite{Nathan2022}. In order to achieve optical amplification with  topological frequency conversion in Weyl semimetals,
electronic relaxation must  be  slow enough to ensure   that  the  energy conversion rate exceeds that of dissipative loss. 
Ref.~\cite{Nathan2022} estimates an 
amplification threshold for the relaxation time $\tau$ at $\sim 100 ~ {\rm ps}$ for realistic material parameters---orders of magnitude larger than the sub-ps timescales typically found in experiments~\cite{Dai_2015,Weber2017}. 
These challenges motivate the search for alternative solid-state platforms for  topological frequency conversion.

\begin{figure}[t]  
\centering
\includegraphics[width=\columnwidth]{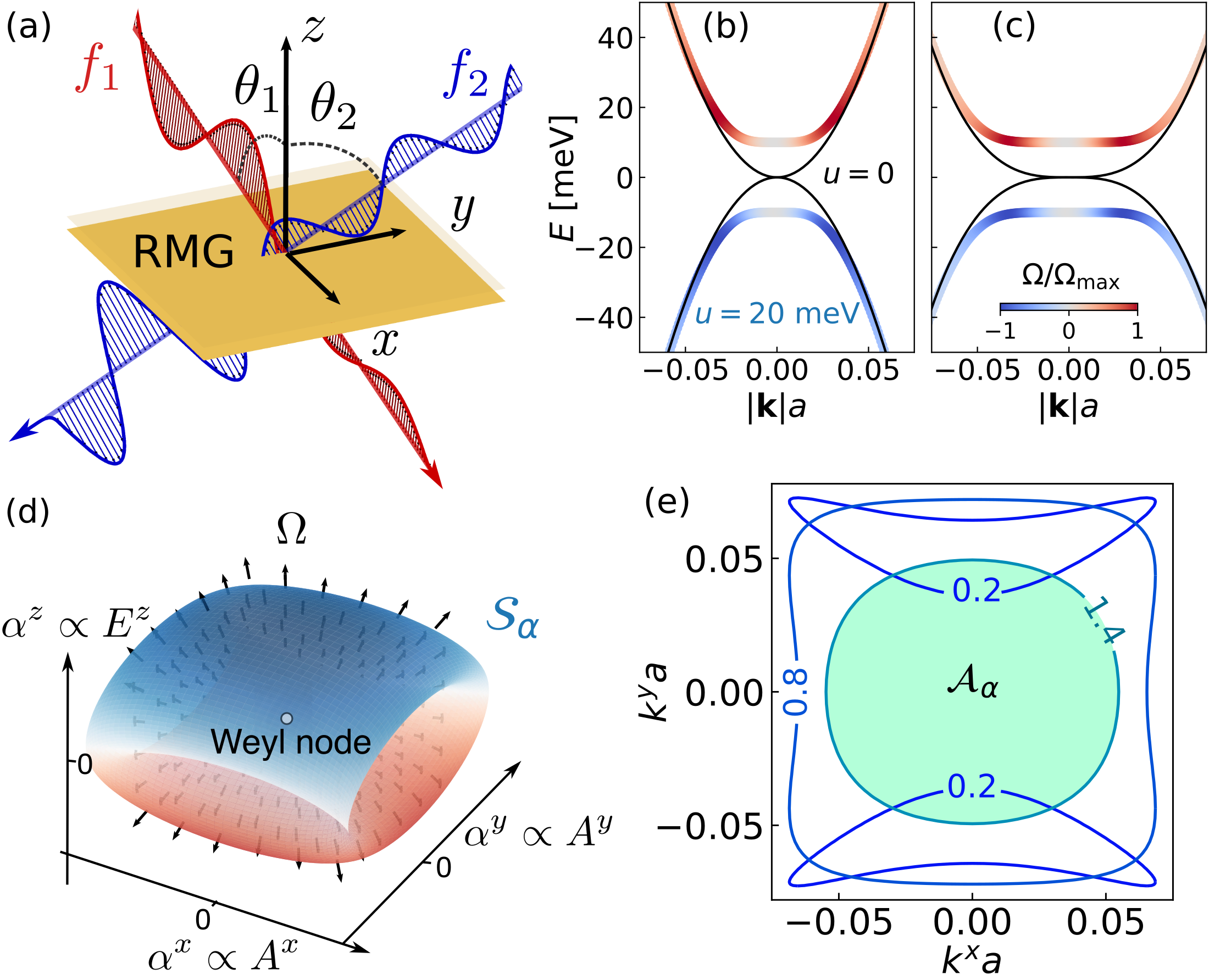}
  \caption{
  (\textbf{a}) Rhombohedral multilayer graphene (RMG) can mediate topological frequency conversion between two incident
  driving modes with frequencies $f_1$, $f_2$. 
 An interlayer potential $u$ gaps out the low-energy bands of RMG [shown for (\textbf{b}) bilayers and (\textbf{c}) trilayers], with a Berry curvature distribution [color scale] controlled by ${\rm sgn}(u)$. (\textbf{d}) The driving modes couple to electrons via their in-plane vector potential and perpendicular electric field, contained in the 3D  vector $\balpha$~[Eq.~\ref{eq:drive_fields_alpha}]. 
  The Berry curvature ${\bm \Omega}$ in $\balpha$ space  (black arrows) hosts a monopole, or Weyl node, located at $(-k^x, -k^y, 0)$ for an electron  with wavevector $(k^x, k^y)$. Under quasi-periodic driving, the system explores a 2D surface ${\cal S}_{\balpha}$, shown here for the drive configuration in \textbf{a}. Frequency conversion power is given by the quantized flux of Berry curvature $\bOmega$ through ${\cal S}_{\balpha}$, see Eq.~\eqref{eq:conversion_power}. (\textbf{e}) The intersection between ${\cal S}_{\balpha}$ and the $E^z = 0$ plane defines a momentum-space area $\cal A_{\balpha}$ where electrons contribute to conversion power at a quantized rate $\pm h f_1 f_2 L$. Contours are labeled by the ratio of electric fields $E^z_0/E_1$ 
  {(see main text) and indicate the corresponding boundaries of $\cal A_{\balpha}$.}
  }
   \label{fig1}
\end{figure}

Here we identify rhombohedral multilayer graphene (RMG)~\cite{Guinea2006, Min2008} as a promising
candidate towards this goal. Similarly to Weyl semimetals, RMG hosts Weyl nodes, but in a  hybrid three-dimensional space formed by the two crystal momenta
and the interlayer potential difference.  This hybrid space can be dynamically addressed via 
the in-plane vector potential and out-of-plane
electric field induced by the driving modes. 
Under suitable configurations of incident beams, a fraction of charge carriers in the system then converts energy between the modes at the quantized rate $h f_1f_2$, as illustrated in Fig.~\ref{fig1}.

The quasi two-dimensional nature of RMG furnishes key advantages for frequency conversion including reduced screening and lack of skin effects, along with gate tunability which provides an extra lever of control. Due to its pristine nature, relaxation rates in RMG are also expected to be much lower than in bulk materials, such as Weyl semimetals. Moreover, as a unique feature of our setup, the different couplings to in-plane and out-of-plane radiation fields lead to a strong sensitivity on the angle and polarization of the incident beams. Within the rhombohedral family, we find that Bernal bilayer graphene appears most promising for THz-scale implementations due to its lower propensity towards dissipation.

The two time-reversed valleys of RMG contribute to topological frequency conversion with an opposite sign; a nonzero net conversion rate hence requires an imbalance in the electronic distributions of the two valleys. We conclude our paper by discussing various  strategies for inducing such valley-asymmetric electronic distributions. 

\emph{\textbf{Setup.}} RMG consists of $L$ graphene sheets stacked such that 
each $A$ sublattice 
lies above the $B$ sublattice of the layer below---these superimposed sites are coupled by the leading interlayer tunneling term $\gamma_1 \approx 360$ meV.
Near charge neutrality, the 
wavefunctions of carriers
primarily reside in the  subspace ($A_1, B_L$) comprising the leftover $A$ sites of the bottom layer and $B$ sites of the top layer. The system can thus be described by an effective two-level Hamiltonian in each of the two valleys~\cite{Min2008}, which 
in the presence of an out-of-plane electric field $E^z$ reads
\begin{align}
    H_{\xi}(\bk) =
    \begin{pmatrix}
    u/2 & - \gamma_L \left(\xi k^x - i k^y \right)^L \\
   - \gamma_L \left(\xi k^x + i k^y \right)^L & -u/2 
    \end{pmatrix} ,
    \label{eq:general_Hamiltonian}
\end{align}
with the wavevector $\kv = (k^x, k^y)$ measured from the hexagonal Brillouin zone corner corresponding to valley $\xi=\pm1$. Here $\gamma_L = (\hbar v_F)^L / \gamma_1^{L-1}$, with $v_{\rm F} = 8.4 \times 10^5$ m/s the Fermi velocity  of graphene, while $u = d (L-1) e E^z$ denotes the potential difference induced by $E^z$, with $d = 0.33$~nm the interlayer distance and $e$ the electron charge. For $u=0$ the low-energy excitations of RMG are 
chiral fermions with an $L$-th power dispersion~\cite{footnote1}.
A nonzero interlayer potential $u$  opens 
a  gap in the dispersion and controls the Berry curvature of the low-energy bands (Fig.~\ref{fig1}b,c)---a feature at the heart of our proposal.

Bichromatic driving induces an electric field $\vec E(t) = \vec E_1(t) + \vec E_2(t+\Delta t)$ in  the system. Here $\vec E_j(t)$ is the monochromatic
field from mode $j$ with frequency $f_j$,
and $\Delta t$ controls the relative phase shift between the modes---relevant   when $f_1$ and $f_2$ are commensurate. Due to the quasi-2D nature of RMG we disregard skin effects 
and plasmon excitations that can become important in 3D implementations such as Weyl semimetals~\cite{Nathan2022}. We further consider THz-scale frequencies,
corresponding to wavelengths $\lambda \sim 100 \, \rm \mu m$, such that the electric field can be considered uniform in the sample. The in-plane electric field component $\vec E^\parallel$ couples to electronic crystal momenta via the Peierls substitution $\bk \rightarrow \bk + e \bA^{\parallel}(t)/\hbar $, with $\vec A^\parallel(t)$ the net vector potential induced by $\vec E^\parallel(t)$.
In contrast, the out-of-plane component
$E^z$ couples through the effective dipole moment in the low-energy subspace, Eq.~\eqref{eq:general_Hamiltonian}. The coupling of radiation 
thus
strongly depends on the incidence angles $\theta_j$ and polarization of the drives (see Fig.~\ref{fig1}a), a unique feature of this setup.

The driven system is conveniently described by the two-level Hamiltonian
$H_{\xi}(\bk, t)=\vec d_\xi(\bk ,t)\cdot {\bm \sigma}$,
with ${\bm \sigma} = \left(\sigma^x, \sigma^y, \sigma^z\right)$ a vector of Pauli matrices acting in the low-energy subspace and 
\begin{equation}
   \vec d_\xi(\bk ,t) = \gamma_L \left( - \text{Re}[ 
   \Pi_{\xi}^L(\bk ,t)], - \text{Im} [\Pi_\xi^L(\bk ,t)] ,
   \alpha^z(t) \right).
   \label{eq:H_twolevel}
\end{equation}
Here $\Pi_{\xi}(\bk ,t) = \xi (k^x + \alpha^x(t)) + i (k^y + \alpha^y(t))$ and $\balpha \equiv \balpha_1+\balpha_2$ encode the
coupling to 
electromagnetic fields,
\begin{equation}
\balpha_j(t) = e \left( \frac{A_j^x(t)}{\hbar}, \frac{A_j^y(t)}{\hbar},  \frac{d (L-1)  E_j^z(t)}{2 \gamma_L} \right).
\label{eq:drive_fields_alpha}
\end{equation}
Below, we find it convenient to {recast the time dependence of } $\balpha_j$ (and therefore $\bd_\xi$) 
in terms of the phases of the modes,
$\phi_j = \omega_j t$ with $\omega_j = 2\pi f_ j$. 

\emph{\textbf{Frequency conversion.}}
To see how topological frequency conversion emerges, 
consider the time-averaged rate of work $P_j^\xi(\kv)$ done by mode $j$ on a valence-band  electron in valley $\xi$ and wavevector $\kv$.
Neglecting dissipative effects for now, energy conservation implies $P_{1}^{\xi} (\kv) = -P_2^\xi(\kv) = P_{\rm FC}^\xi(\bk)$, 
with the {\it frequency conversion power} $P_{\rm FC}^\xi(\bk)$ giving the time-averaged 
 energy transfer rate from mode $1$ to mode $2$ mediated by the electron. 

When the time dependence of  $H_\xi(\kv,t)$ is quasi-adiabatic, $P_{\rm FC}^\xi(\bk)$
is proportional to 
the number of times the surface spanned by the 
unit vector $ \hat{\bd}_\xi \equiv \vec{\bd_\xi}/|\vec{\bd_\xi}|$
encloses the origin~\cite{Martin2017,Nathan2022, Esin2023},
\begin{equation}
P_{\rm FC}^{\xi}(\bk) = -h f_1 f_2 \int_0^{2\pi} \frac{d\phi_1d\phi_2 }{4 \pi} \hat{ \vec d}_{\xi}\cdot (\partial_{\phi_1}\hat{\vec d}_{\xi}\times \partial_{\phi_2}\hat{\vec d}_{\xi})  .  
\end{equation}
This result is derived in the Appendix and emerges from the leading-order diabatic correction to the dynamics---along with the fact that, for incommensurate frequencies, the system uniformly explores the two-dimensional parameter space $0 \leq \phi_j \leq 2\pi$.
For commensurate frequencies, this formula is recovered by also averaging $P_{\rm FC}^\xi(\kv)$ over $\Delta t$.
Changing variables from $\phi_j$ to $\balpha$ yields
\begin{equation}
P_{\rm FC}^\xi(\kv) = - \frac{ h f_1 f_2}{2\pi} \int_{{\cal S_{\balpha}}}  {d\vec S_{\balpha}} \cdot {\bm \Omega_\xi}(\kv,\balpha) ,
\label{eq:conversion_power}
\end{equation}
with $\int_{\mathcal S_{\balpha}} d\vec S_{\balpha} = \int  d\phi_1 d\phi_2 \partial_{\phi_1} \balpha \times \partial_{\phi_2} \balpha $ the oriented integral on the surface
$\mathcal S_{\balpha}$ spanned by $\balpha(\phi_1, \phi_2)$, 
and $\vec \Omega_\xi$ the vector-valued  Berry curvature associated with the winding of
$\bd_\xi$, $\Omega^i_\xi = \frac{\epsilon_{ijl}}{2} \hat{\vec d}_\xi \cdot (\partial_{\alpha^j} \hat{\vec d}_\xi \times \partial_{\alpha^l}\hat{ \vec d}_\xi)$.
Thus, {frequency conversion power} is controlled by the Berry curvature flux through 
${\cal S}_{\balpha}$ (see Fig.~\ref{fig1}d). The overall sign in Eq.~\ref{eq:conversion_power} is reversed for conduction band electrons.

In analogy to Weyl semimetals, sources and sinks of Berry curvature in the hybrid three-dimensional space $\balpha$ are quantized and {confined to} topologically-protected gap-closing points, analogous to Weyl nodes~\footnote{The term Weyl node  normally refers to nodal points in the 3D parameter space comprising the crystal momenta in bulk materials. Here we use the term ``Weyl node'' in a generalized sense to describe gap closings as a function of any three parameters.}. Indeed,
the RMG Hamiltonian in Eq.~\eqref{eq:H_twolevel} hosts a charge-$L$ Weyl node at $\balpha_0 (\kv)= (-k^x, -k^y, 0)$, 
\begin{equation}
\nabla_{\balpha} \cdot {\bOmega_\xi} (\kv,\balpha)= 2 \pi \xi L \delta(\balpha-\balpha_0(\kv)).
\label{eq:monopole}
\end{equation}
A valence band electron with wavevector $\kv = (k^x, k^y)$ therefore acts as a topological frequency converter if $\balpha_0(\kv)$ is located within the surface $\cal S_{\balpha}$. This condition defines an area $\cal A_{\balpha}$ in the Brillouin zone (see Fig.~\ref{fig1}e) where each valence-band electron with
momentum $ \kv\in \cal A_{\balpha}$ 
mediates energy transfer at the quantized rate $P_{\rm FC}^\xi(\kv) = \pm  h f_1 f_2 \xi L$. Here, the  $+$ and $-$ signs
occur when the area element $d \vec S_{\balpha}$ defined from the driving protocol [below Eq.~\eqref{eq:conversion_power}] points inwards and outwards, respectively.
For electrons outside of $\cal A_{\balpha}$, $P_{\rm FC}^\xi(\kv) = 0$.

We now consider the dynamics of the many-body state. In the regime of  adiabatic driving and slow relaxation relative to the driving frequencies (that is, $\omega_j \tau \gg 1$), the steady-state population of band $n = 0, 1$ near valley $\xi$ takes an approximately stationary value,
    $\bar \rho_n^\xi(\kv)$~\footnote{In a simple relaxation-time approximation, $\bar \rho_n^\xi(\kv)$ is given by the time-averaged value of $f_\beta([\varepsilon_n(\kv,t)-\mu])$ with $f_\beta(E)$ the Fermi-Dirac distribution~\cite{Nathan2022}; we expect more general dissipation mechanisms will lead to qualitatively similar distributions.}.
The frequency conversion intensity (i.e. the conversion power per unit area) from valley $\xi$ is obtained by
weighing the individual contribution $P_{\rm FC}^\xi(\kv)$ of each electron with the corresponding steady-state occupation,
\begin{equation}
\cI_{\rm FC}^\xi = \pm h f_1 f_2 L \xi \int_{\cal A_{\balpha}} \frac{{\rm d}^2 \bk}{(2\pi)^2} \left( \bar \rho_0^\xi(\bk) - \bar \rho_1^\xi(\bk)  \right) .
\label{eq:ConversionRate}
\end{equation}
The boundary of $\cal A_{\balpha}$ consists of the wavevectors $\kv$ for which the instantaneous gap of $H_\xi(\kv,t)$ closes at some time $t$. States near this  boundary thus
undergo inter-band Landau-Zener tunneling, which causes significant dissipation through \emph{Landau-Zener absorption}---whereby electrons  pumped  to the valence  band 
relax dissipatively, leading to an irreversible loss of the invested energy.

To estimate the conversion rate from a single valley at charge neutrality, we set $\bar\rho_0^\xi \approx 1$, $\bar \rho_1^\xi \approx 0$ and approximate $\mathcal A_{\balpha}$ as a rectangle of sides $2 e A_j \cos \theta_j/\hbar$, 
\begin{equation}
    \cI_{\rm FC}^{\xi} \approx {\xi} L \frac{e^2 E_1 E_2 }{2 \pi^3 \hbar} \cos \theta_1 \cos \theta_2,
\end{equation} 
where we used $ A_j \sim E_j/\omega_j$. For electric fields $E_j \sim 0.1$ V/nm and $\theta_j \sim 89^\circ$, $\cI_{\rm FC}^\xi \sim 10 L$ W/mm$^2$ (see also numerics in Fig.~\ref{fig3}). 
Interestingly, the conversion intensity is proportional to the geometric mean of radiation intensities, $I =c \epsilon_0 E_1 E_2/2$, through $\cI_{\rm FC}^\xi/I \approx 4 L  \alpha_{\rm EM}\cos^2( \theta) / \pi^2 $ with $\alpha_{\rm EM}$
the fine-structure constant. 

When the chemical potential $\mu_\xi$ in a given valley moves away from charge neutrality, 
we expect the conversion intensity $I_{\rm FC}^\xi$
to decrease,
due to cancellations between conduction and valence-band carriers (for $\mu_\xi>0$) or de-populating the valence band (for $\mu_\xi<0$). Because of the opposite contributions from  
carriers in valleys $\xi = \pm 1$ (a consequence of time-reversal symmetry), a net effect requires
a valley imbalance in the number of active frequency converters, e.g. through unequal chemical potentials, $\mu_+ \neq \mu_-$, or a valley-selective population inversion.
We discuss strategies to induce such imbalances below.

\emph{\textbf{Numerical simulations.}} We support our conclusions with simulations of Bernal bilayer graphene subjected to bichromatic radiation, including trigonal warping corrections to its low-energy band structure~\cite{Jung2014}. Following Ref.~\cite{Nathan2022}, we  consider the master equation
\begin{equation}
    \partial_t \hat \rho_\xi (\bk, t) = - \frac{i}{\hbar} [\hat H_\xi (\bk,t), \hat \rho_\xi(\bk,t)] + {\cal D}_{\xi}(\kv,t) \left\{ \hat \rho_\xi (\bk, t) \right\} ,
    \label{eq:master_equation}
\end{equation}
where $\hat H_\xi (\bk,t)$ and $\hat\rho_\xi(\bk,t)$ denote the second-quantized Hamiltonian and density matrix of the system.
We model dissipation via 
${\cal D}_{\xi}(\kv,t) \left\{ \hat \rho_\xi (\bk, t) \right\} 
= -\frac{1}{\tau} (\hat \rho_\xi(\bk, t) 
-\hat \rho_{\xi}^{\rm eq}(\bk, t) )$, 
characterizing uniform relaxation on a timescale $\tau$ towards an instantaneous 
equilibrium state with a bath at inverse temperature $\beta$,
$\hat \rho_{\xi}^{\rm eq}(\bk, t) = e^{-\beta [ \hat H_\xi(\bk, t)-\mu_\xi \hat n]} / {\rm Tr}[e^{-\beta [ \hat H_\xi(\bk, t)-\mu_\xi \hat n]}]$,
with $\hat n$ the electron density operator.
For simplicity we work at charge neutrality, $\mu_\xi = 0$; we consider effects of driving in the presence of a Fermi surface (for $\mu_\xi \neq 0$) in the Appendix.

\begin{figure}[t]
	\includegraphics[width=\columnwidth]{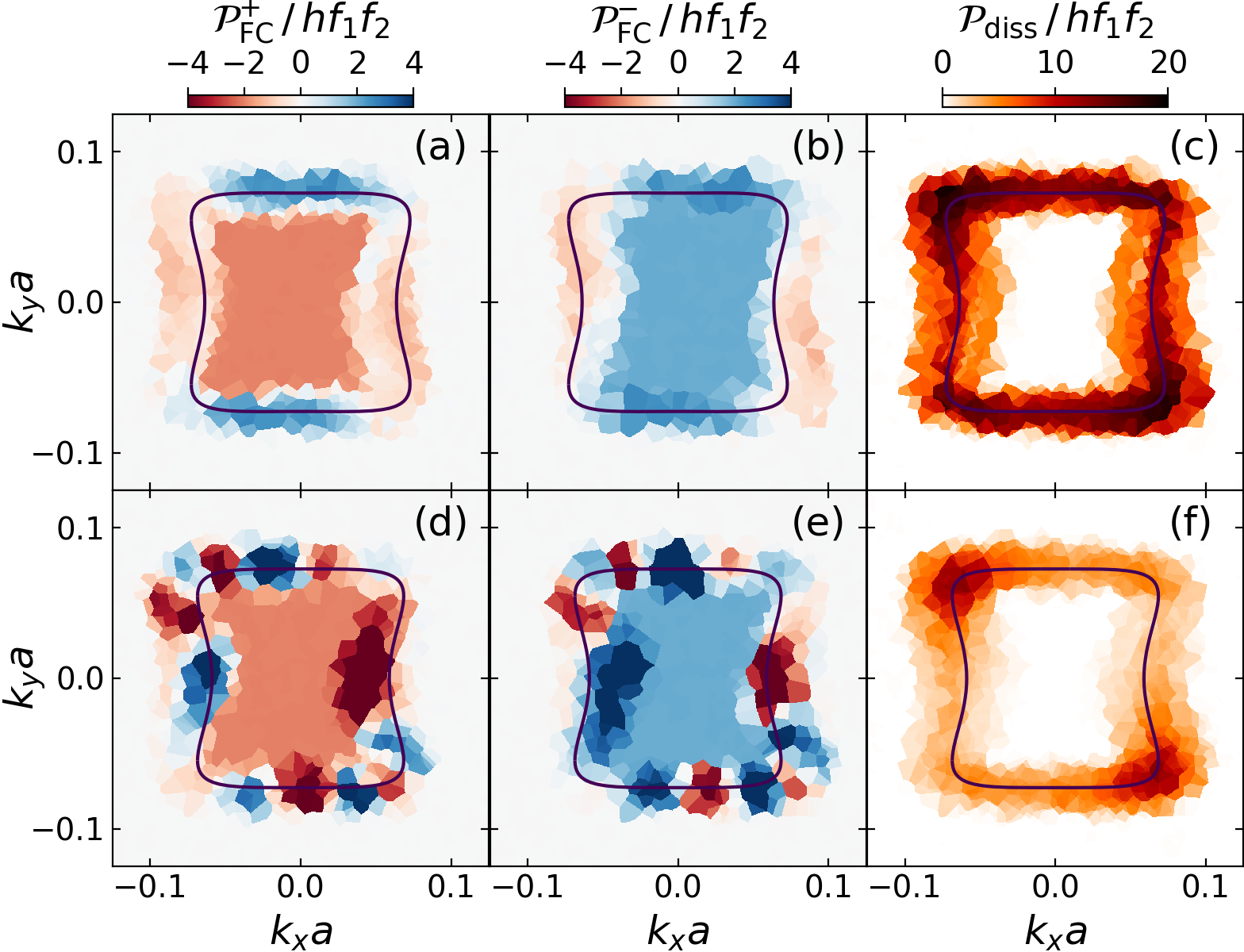}
	\caption{ Momentum-resolved frequency conversion in Bernal bilayer graphene. We consider two linearly-polarized modes
    in the configuration of Fig.~\ref{fig1}a with $\left( \pi/2 - \theta_j \right) \approx 0.1^\circ$, {electric field scale $E_1 = 0.1$ V/nm,} relaxation time $\tau = 500$ ps, and at charge neutrality, $\mu=0$. Contributions to frequency conversion from valleys $\xi = +1$ (\textbf{a},\textbf{d}) and $-1$ (\textbf{b},\textbf{e}) are shown alongside dissipated power (\textbf{c},\textbf{f}).  Frequencies $f_1 \approx 0.128$ THz and $f_2 =f_1/\sqrt{2}$ [(\textbf{a})--(\textbf{c})] or $f_2 = 2f_1/3$ [(\textbf{d})--(\textbf{f})] illustrate quasi-periodic and periodic driving.
    The color scales represent time-averaged powers in units of $h f_1 f_2$. The overlaid contour shows the predicted boundary of ${\cal A}_{\balpha}$.
    }
	\label{fig2}
\end{figure}

We numerically obtain the
steady-state solution to Eq.~\eqref{eq:master_equation} for a representative grid of $\kv$ points and extract the time-averaged power per area transferred into the system from mode $j$ 
through ${\cal I}_j^\xi = \int\frac{d^2\kv}{(2\pi)^2} \cP_{j}^\xi(\kv)$,
where
\begin{equation}
     {\cal P}_{j}^\xi(\bk) = \lim_{T \rightarrow \infty} \int_0^T \frac{dt}{T} ~ \partial_t \balpha_j(t) \cdot {\rm Tr} \left[\hat \rho_\xi(\bk, t)\nabla_{\balpha_j}  \hat H_\xi (\bk, t)  \right] .
\end{equation}
This expression
can be derived from first principles and,  
being agnostic to our theoretical analysis above, serves as an independent check of its conclusions.

We consider two linearly-polarized modes propagating with  incidence angles $\theta_j$ (relative to $\hat{\mathbf{z}}$) in the $xz$ and $yz$ planes, as depicted in Fig.~\ref{fig1}a.
The induced vector potentials are given by
$ \bA_1(t)=  A_1 (\cos \theta_1 \cos [\omega_1t], 0, \sin \theta_1 \cos [\omega_1t])$ and 
$\bA_2(t) =  A_2 (0, -\cos \theta_2 \cos  [\omega_2(t{+\Delta t})], \sin \theta_2 \cos[\omega_2(t{+\Delta t})])$,
with $\Delta t$ controlling the phase shift between the modes. We consider large angles of incidence $\theta \approx \pi/2$, to compensate for the  weaker effective coupling between the radiation fields and $\sigma^z$ in Eq.~\ref{eq:H_twolevel} (as compared to the in-plane components of the radiation). In order to obtain a finite area $\cal A_{\balpha}$ we include a time-independent component of the out-of-plane field (coming from, e.g., a substrate or a back gate), such that $E^z = -\partial_t A^z + E^z_0$ (see Fig.~\ref{fig1}d,e). In our simulations, we consider $E_1 = 0.1$ V/nm and $E^z_0 \approx 2/3 E_1$, and set $E_2 = (\omega_2/\omega_1) E_1$ so that the conversion region $\cal A_{\balpha}$ remains unchanged for different frequency ratios. This electric field scale induces momentum displacements
within the region of validity of the low-energy description, $e A_j \cos \theta_j  \lesssim \gamma_1/2v_F$~\cite{footnote1}.

\begin{figure}[t]
   	\includegraphics[width=\columnwidth]{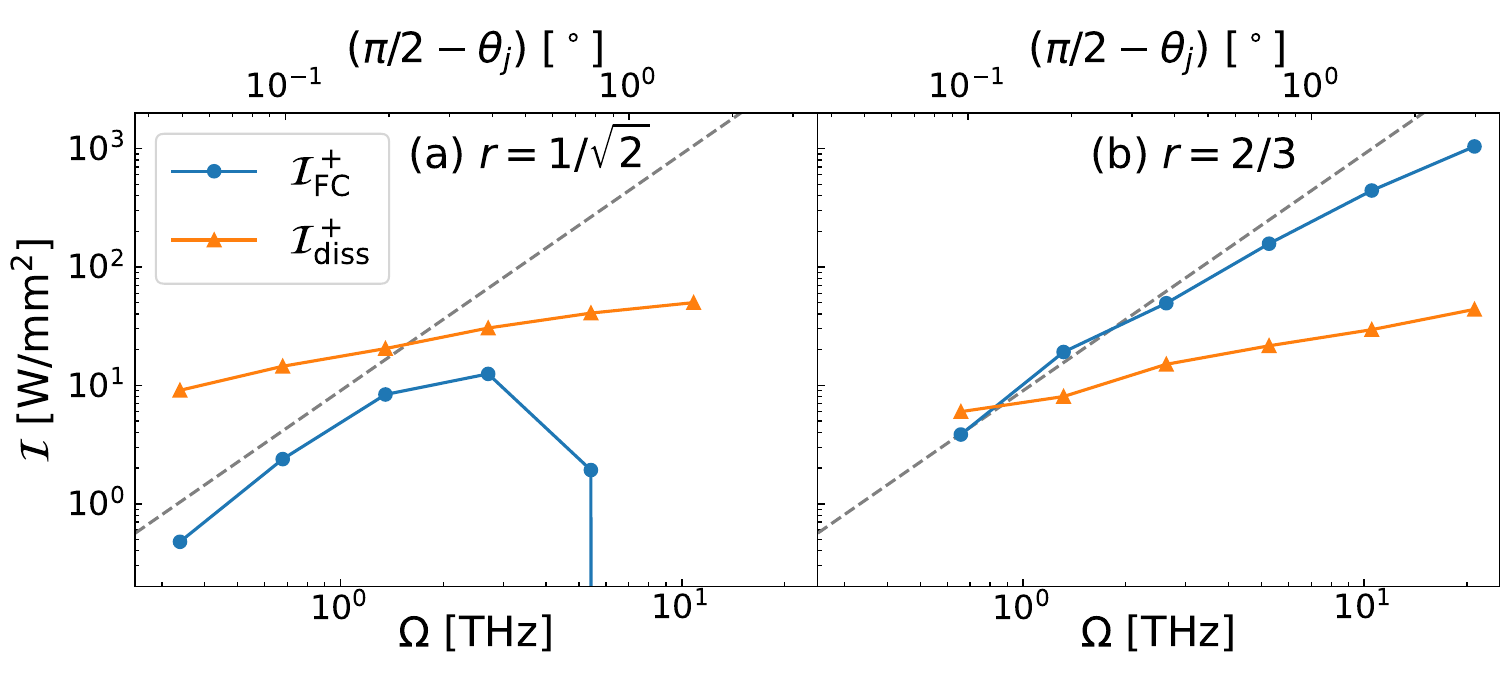}
	\caption{Valley-resolved frequency conversion ${\cal I}_{\rm FC}^+$ and dissipation ${\cal I}_{\rm diss}^+$ intensities in Bernal bilayer graphene
    as a function of $\Omega = \sqrt{\omega_1 \omega_2}$. The incidence angles $\theta_j$ are tuned simultaneously with $\Omega$
    to keep the in-plane vector potential amplitude ($A_j \propto \cos \theta_j/\omega_j$) constant, using Fig.~\ref{fig2} as a reference point.
    The dashed lines indicate the scaling law $\cI_{\rm FC}^+ \sim \Omega^2$, which describes the low-frequency limit where non-adiabatic corrections can be neglected. The two panels contrast (\textbf{a}) incommensurate and (\textbf{b}) commensurate frequency ratios $r=1/\sqrt{2}$ and $2/3$, respectively. For commensurate drives, frequency conversion overpowers dissipative losses in the THz regime. Non-adiabatic effects cause the deviation of $\cI_{\rm FC}^+$ from quadratic scaling at high frequencies; they are more prominent for incommensurate driving, leading to a vanishing conversion power for $\Omega\gtrsim 5\,{\rm THz}$ in (\textbf{a}).} 
	\label{fig3}
\end{figure}

We first set $\omega_1 \approx 0.8$ THz and probe different frequency ratios $r=\omega_2/\omega_1 $.
Fig.~\ref{fig2} shows the valley-resolved conversion power $\cP_{\rm FC}^{\pm}(\kv) = \frac{1}{2} ( \cP_{\rm 1}^\pm(\kv) - \cP_{2}^\pm(\kv) )$, obtained for $r =1/\sqrt{2}$ (a--b) and $r={2}/{3}$ (d--e), respectively.
For the latter case, we compute $\cP_{\rm FC}^{\pm}(\kv)$ by averaging over $\Delta t$.
Consistent with our analysis, 
electrons located within ${\cal A}_{\balpha}$ transfer energy between the modes at the quantized rate $\pm 2 h f_1 f_2$. 
Near the boundary, these electrons generate substantial dissipation, $\cP_{\rm diss}(\bk) =  \sum_\xi ( \cP_{\rm 1}^\xi(\kv) + \cP_{2}^\xi(\kv) )$, through the Landau-Zener absorption mechanism [Fig.~\ref{fig2} (c,f)].
Dissipation is strongly reduced for commensurate driving due to the mechanism of Lissajous conversion~\cite{Nathan2022}: 
Only a few electrons on the boundary of $\mathcal A_{\balpha}$ are taken to the Weyl node by $e\vec A(t)/\hbar$ and hence
generate Landau-Zener absorption. The anisotropic profile of conversion power along the boundary
arises from the distinct mode polarizations, which lead to a significant $\bk$ dependence on the mode most responsible for pumping electrons between bands.

\begin{figure}[t]
    \includegraphics[width=\columnwidth]{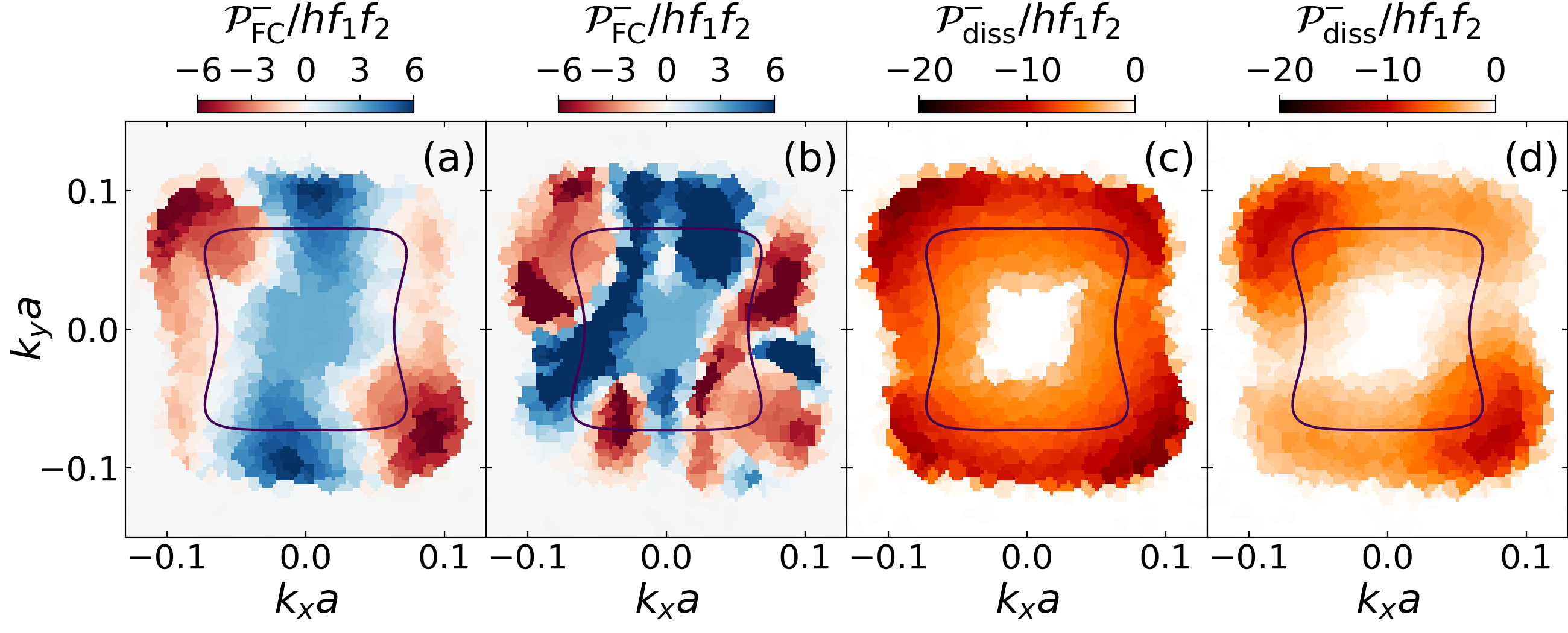}
	\caption{Frequency conversion power (\textbf{a},\textbf{b}) and dissipation (\textbf{c},\textbf{d}) in valley $\xi=+1$ of rhombohedral trilayer graphene. We consider quasi-periodic (\textbf{a},\textbf{c}) and periodic (\textbf{b},\textbf{d}) driving, with parameters identical to Fig.~\ref{fig2}. While electrons in the quantized regions contribute $50\%$ more to frequency conversion than in bilayer graphene, due to Weyl nodes carrying charge $L=3$, dissipation through Landau-Zener absorption is also enhanced because of the flatter low-energy dispersion.}
	\label{fig4}
\end{figure}

In Fig.~\ref{fig3} we plot the integrated conversion intensity from a single valley, 
${\cal I}_{\rm FC}^+ \equiv  \frac{1}{2} \left({\cal I}_1^+ - {\cal I}_2^+ \right)$, and the corresponding dissipated intensity ${\cal I}_{\rm diss}^+\equiv  ({\cal I}_1^+ + {\cal I}_2^+)$ 
as a function of frequency, with the same values of $r$ as in Fig.~\ref{fig2}. 
To remove geometric factors from the analysis, 
we scale the incidence angle $\theta_j$ 
such that the in-plane vector potential amplitude, $A^{\parallel}_j \propto E_ j \cos \theta_j / \omega_j$, remains constant---thus
the surface $\cal S_{\balpha}$ is unchanged and only the time-dependence of $\balpha(t)$ is rescaled. 
In both commensurate and incommensurate cases, {$\cI_{\rm FC}^+$  
grows quadratically with $\Omega$ at low frequencies}, following the scaling of the conversion power quantum $h f_1 f_2 L$. At higher frequencies, the expansion of the region with strong Landau-Zener tunneling, near the boundary 
of $\cal A_{\balpha}$, leads to a sub-quadratic frequency scaling of $\cI_{\rm FC}^+$ (see also Figs.~\ref{fig5} and \ref{fig6} in the Appendix). These non-adiabatic effects are much stronger for incommensurate driving protocols [Fig.~\ref{fig3}(a)]; beyond a critical frequency they lead to a sharp decrease in conversion power.
Crucially, for commensurate driving on the THz scale [Fig.~\ref{fig3}(b)], $\cI_{\rm FC}^\xi$ can vastly outperform dissipative losses with $\tau$ of order $100 - 1000$ ps. In the Appendix we also estimate typical heating rates of order $0.1 - 1$ K/ps, potentially allowing to perform topological frequency conversion over many THz drive cycles.

To demonstrate the extension
of topological frequency conversion to other members of the RMG family, we simulate rhombohedral trilayer graphene irradiated by the same bichromatic driving as above (Fig.~\ref{fig4}). 
Our results show quantized plateaus hosting topological frequency conversion at the rate $\pm 3 h f_1 f_2$ per electron (due to the charge-$3$ Weyl nodes in this system). This
gain compared to bilayer graphene is however offset by a larger dissipation rate, due to the flatter low-energy dispersion that enhances the non-adiabatic regions of momentum space---a trend expected to continue with increasing number of layers~\cite{Guinea2006, Min2008}.

\emph{\textbf{Outlook.}} We have shown that
RMG can 
mediate energy transfer between
electromagnetic modes through the mechanism of  topological frequency conversion. 
The two-dimensional nature of this platform results in 
an anisotropic coupling to radiation, which in turn leads to a strong dependence of conversion power on the incidence angle and polarization of the drives.

Our results indicate that amplification from a single valley can be achieved with commensurate frequencies in the THz range  and relaxation timescales $\tau$ of order $100-1000$ ps.
In order to achieve a net conversion, however, time-reversal symmetry (which relates 
the two valleys) must be broken. We speculate that valley-imbalanced populations can be realized either spontaneously
through electronic interactions at low temperatures~\cite{Jung2015, Shi2020, Zhou2021a, Kerelsky2021, Zhou2021b, Zhou2022, delaBarrera2022, Seiler2022,  liu2023interactiondriven, Zhang2023, Han2023, Han2023b, Sha2024, Han2024}, by interfacing with magnetic/spin-orbit coupled materials~\cite{Vila2021, Zhang2023, Ming2023, Han2023, Han2023b, Sha2024, Han2024, Wang2024, Zhumagulov2024, Koh2024, Zhumagulov2024b, Koh2024b}, or through photoexcitations generated by circularly-polarized light~\cite{Abergel2010, Friedlan2021, Kumar2021, Yin2022, kumari2024valley}. Recent experiments suggest that valley-imbalanced transient states can be long-lived (with, e.g.,  valley lifetimes reported in a large range from $1 \mu$s to $100$ms in bilayer graphene quantum dot experiments with applied magnetic fields~\cite{Garreis2024, banszerus2024phononlimited}). The investigation of such schemes present an interesting opportunity for further studies of frequency conversion phenomena. 

Among the RMG family, we find Bernal bilayer graphene to be most promising for achieving THz-scale amplification, due to its steeper low-energy dispersion which minimizes dissipation from Landau-Zener absorption. With low-temperature mean free paths of order $l \sim 10 ~ \mu{\rm m}$~\cite{Zhou2022,Seiler2022, delaBarrera2022} and typical Fermi velocities $v \sim 10^5$ m/s
near charge neutrality, electronic relaxation timescales $\tau \sim l/v \sim 100$ ps appear within experimental reach. Provided that valley-imbalanced populations can be achieved and controlled, our simulations indicate that THz frequency conversion can overpower dissipative loss in this regime and occur before excessive heating sets on. 

Finally, topological frequency conversion in RMG requires relatively strong electric fields and precise alignment of the polarization of the incoming radiation. We speculate that these challenges can be addressed by
focusing the beams
using waveguides and antennas---such structures, along with backgates of suitable geometry providing the required constant electric fields, could become part of a future frequency conversion device.

\begin{acknowledgments}

\emph{\textbf{Acknowledgments}.}
We thank Cyprian Lewandowski and Christopher Yang for insightful discussions {and collaborations on related projects}. \'E. L.-H. was supported by the Gordon and Betty Moore
Foundation’s EPiQS Initiative, Grant GBMF8682.
F.N. was supported by  the U.S. Department of Energy, Office of Science, Basic Energy Sciences under award DE-SC0019166, the Simons Foundation under award 623768, and the Carlsberg Foundation, grant CF22-0727.
G.R. and I.E. are grateful for support from the Simons
Foundation and the Institute of Quantum Information
and Matter. 
G.R. is grateful for support from the ARO MURI grant FA9550-22-1-0339. 
This work was performed in part at Aspen Center for Physics, which is supported by National Science Foundation grant PHY-2210452.
\end{acknowledgments}

\bibliography{biblio}

\begin{thebibliography}{71}
\bibitem [{foo()}]{footnote1}%
  \BibitemOpen
  \href@noop {} {\bibinfo {title} {For $|\bk| \gtrsim \gamma_1/2 \hbar v_f$,
  the dispersion crosses over to linearly-dispersing dirac cones and the
  eigenstates have support beyond the sublattices $a_1$, $b_l$.}}\BibitemShut
  {Stop}%
\end{thebibliography}

\clearpage

\appendix

\setcounter{figure}{0}
\renewcommand{\figurename}{Fig.}
\renewcommand{\thefigure}{S\arabic{figure}}

\begin{widetext}

\section{SUPPLEMENTARY MATERIAL \\
{\normalfont for} \\
\vspace{0.3cm}
Topological frequency conversion in rhombohedral graphene multilayers \\
{\normalfont \'Etienne Lantagne-Hurtubise, Iliya Esin, Gil Refael,
and Frederik Nathan} 
}
\end{widetext}

\section{Topological frequency conversion in slowly-driven two-level systems}\label{app:single_electron_conversion}

In this Appendix we review the physical mechanism of topological frequency conversion, following the approach in Ref.~\cite{Nathan2022}, and its application to rhombohedral multilayer graphene. We consider a single valence-band electron at crystal momentum $\bk$ and in valley $\xi$, and omit these indices below for notational simplicity.

We start with the general expression for the instantaneous rate of work $P(t)$ done on the system from the driving fields, 
\begin{equation}
P(t)=  \langle \psi(t)|\partial_t H( t) |\psi(t)\rangle,
\end{equation}
where $H(t) = \bd(t) \cdot \bsigma$ is the driven two-level Hamiltonian parametrized by Eq.~\ref{eq:H_twolevel} and $|\psi(t)\rangle$ is the instantaneous state of the electron. 
As discussed below, $|\psi(t)\rangle$ differs slightly from the instantaneous ground state of $H(t)$ due to diabatic corrections. Expressing the corresponding Bloch vector as 
$\vec v(t) = \langle \psi(t) | \bsigma | \psi(t) \rangle$, we have
\begin{equation}
P(t)=  \vec v(t) \cdot \partial_t \vec d(t).
\label{eqa:p_def}\end{equation}

We now rewrite the time-dependent Hamiltonian Bloch vector $\vec d(t)$ as a function of the phases of the two modes, $\phi_1 =\omega_1 t$ and $\phi_2 = \omega_2[t+\Delta t]$, such that 
$\vec d(t) = \bd (\omega_1 t,\omega_2 t+\Delta \phi)$, with the phase difference $\Delta \phi \equiv \omega_2 \Delta t$. Then
\begin{equation}
\partial_t \bd(t) = [\omega_1 \partial_{\phi_1}+\omega_2 \partial_{\phi_2}]\bd(\omega_1 t ,\omega_2 t+ \Delta \phi).
\end{equation}
Inserting the above into Eq.~\eqref{eqa:p_def}, we recognize  the contribution $\vec v \cdot \omega_j\partial_{\phi_j}\bd$ as the rate of work done by mode $j$, $P_j$.
The time-averaged conversion power from mode $1$ to mode $2$ is given by
$P_{\rm FC} =  P_1 = -P_2$ (neglecting dissipation), such that
\begin{equation}
P_{\rm FC} = \omega_1 \lim_{T\to \infty}\int_0^T \frac{dt}{T} \vec v(t) \cdot \partial_{\phi_1}\bd(t) .
\label{eq:PFC_appendix}
\end{equation}
To evaluate the integrand, we note that for quasi-adiabatic driving {and slow relaxation,} the time-dependent Bloch vector of the system $\bv(t)$
{can be expressed as
\begin{equation}
\vec v = -\bdh + \frac{\hbar}{2|\bd|} \bdh \times \partial_t \bdh
+ \mathcal O\left(\frac{\omega^2}{|\bd ^2|}\right),
\label{eq:steady_state_result}
\end{equation}
where the second term on the r.h.s. denotes the leading-order diabatic correction and is ultimately responsible for topological frequency conversion.} 
The result Eq.~\ref{eq:steady_state_result} can be derived from a sequence of rotating frame transformations, see e.g. Refs.~\cite{Berry_1987,Esin2023}. 
We neglect $\mathcal O(\omega^2/|\vec d|^2)$ subleading nonadiabatic corrections in the following. 
In this case, we can express $\vec v$ explicitly in terms of the phases $\phi_1$ and $\phi_2$,
\begin{equation}
\vec v \approx - \bdh + \frac{\hbar}{2|\bd|}\bdh\times [\omega_1 \partial_{\phi_1} + \omega_2 \partial_{\phi_2}]\bdh .
\end{equation}

Plugging this result into Eq.~(\ref{eq:PFC_appendix}) and  using  $\partial_{\phi_i}\bd = \bdh \partial_{\phi_i}|\bd| +|\bd |\partial_{\phi_i}\bdh$, along with $\bdh \cdot \partial_{\phi_i} \bdh = \frac{1}{2} \partial_{\phi_i}(\bdh\cdot \bdh) = 0$, {we find 
\begin{equation}
\bv \cdot \partial_{\phi_1} \bd \approx  -\partial_{\phi_1}|\bd| 
+ \frac{\hbar}{2} \left( \bdh\times [\omega_1 \partial_{\phi_1} + \omega_2 \partial_{\phi_2}]\bdh \right) \cdot \partial_{\phi_1}\bdh 
\end{equation}
where we also used that $( \bdh \times \partial_i \bdh )\cdot \bdh = 0$. 
Next, we use the identity $ (\vec a \times \vec b) \cdot \vec c = (\vec c \times \vec a) \cdot \vec b$ along with the antisymmetry of the cross product to {find 
\begin{align}
\bv \cdot \partial_{\phi_1} \bd = -  \partial_{\phi_1}|\bd| 
- \frac{\hbar\omega_2 }{2}\bdh \cdot (\partial_{{\phi_1}}\bdh\times \partial_{{\phi_2}}\bdh) .
\label{eq:appendix_result}
\end{align}

We compute $P_{\rm FC}$ by averaging Eq.~\eqref{eq:appendix_result} over time. 
To this end, we  restore explicit time (or phase) dependence
and note that $\vec v(\phi_1,\phi_2)\cdot \partial_{\phi_1}\bd (\phi_1,\phi_2)$ is $2\pi$-periodic in its arguments. 
For incommensurate frequencies, time-averaging is thus equivalent to phase-averaging:
\begin{align}
    &\lim_{T\to \infty}\int_0^T \frac{dt}{T} [\vec v\cdot\partial_{\phi_1} \bd](\omega_1t,\omega_2 t+ \Delta \phi)\notag \\ &=\int_0^{2\pi}\frac{d^2\phi}{4\pi^2}[\vec v\cdot \partial_{\phi_1}\bd](\phi_1,\phi_2).
\end{align}
For commensurate frequencies,  $P(\omega_1 t ,\omega_2 t + \Delta \phi)$ is not identical to phase-averaging because the system only explores a one-dimensional, closed contour on  the ``phase 
Brillouin zone'' spanned by $\phi_1, \phi_2$.
However, we recover the incommensurate driving result when also averaging $P_{\rm FC}$ over the phase difference $\Delta \phi$ between the two modes. This procedure was used in our numerical simulations for commensurate driving presented in the main text.

Finally, using that $\int d^2\phi\,  \partial_{\phi_1} |\bd |=0$ due to periodic boundary conditions, we obtain the final result for the time-averaged (and phase-averaged for commensurate drives) frequency conversion power,
\begin{align}
P_{\rm FC} \approx  -
\frac{\hbar \omega_1\omega_2}{2 \pi}\int_0^{2\pi} \frac{{\rm d}\phi_1 \phi_2}{4 \pi} \bdh \cdot (\partial_{\phi_1}\bdh\times \partial_{\phi_2}\bdh).
\end{align}
Using $\frac{\hbar \omega_1\omega_2}{2\pi} =  hf_1f_2$, we obtain the result quoted in Eq.~\ref{eq:conversion_power} of the main text.

\section{Scaling of conversion power and dissipation with frequency}

\begin{figure*}[t]
	\includegraphics[width=0.98\textwidth]{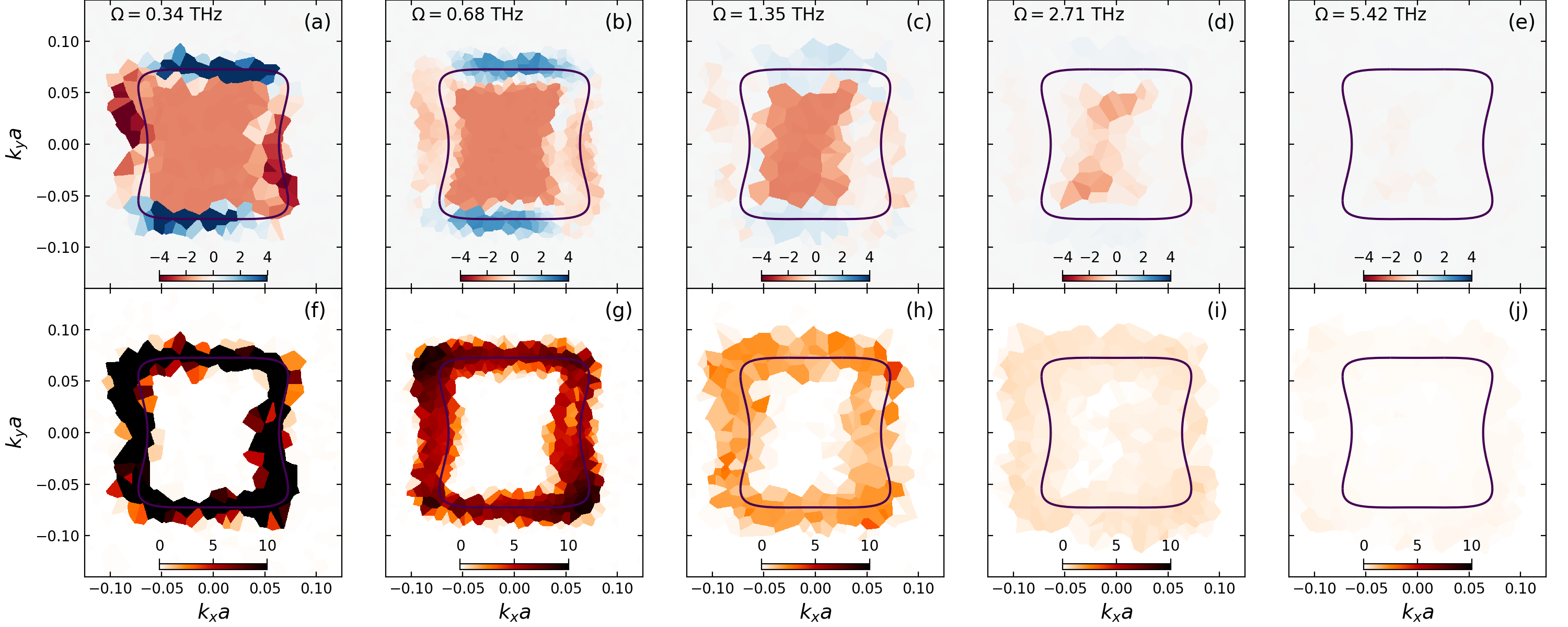}
	\caption{Scaling of the (time-averaged) frequency conversion power $\cP^+_{\rm FC}$ (panels a--e), and the corresponding dissipated power  $\cP^+_{\rm diss}$ (panels f--j), as a function of the geometric mean of incident frequencies, $\Omega = \sqrt{\omega_1 \omega_2}$ (increasing from left to right). Color bars are expressed in units of the power quantum $h f_1 f_2$. We consider incommensurate drives with $r=1/\sqrt{2}$.}
	\label{fig5}
\end{figure*}

\begin{figure*}[t]
	\includegraphics[width=0.98\textwidth]{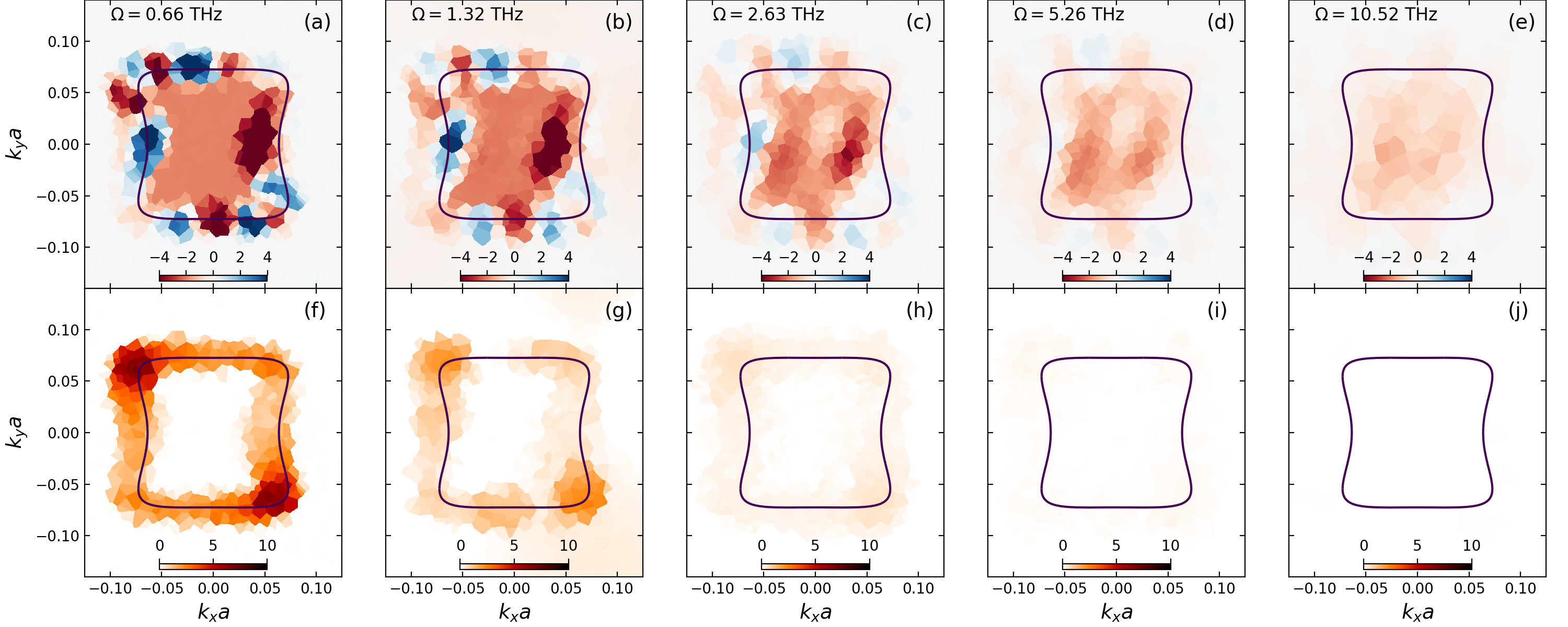}
	\caption{Scaling of the (time-averaged) frequency conversion power  $\cP^+_{\rm FC}$ (panels a--e), and the corresponding dissipated power  $\cP^+_{\rm diss}$ (panels f--j), as a function of the geometric mean of incident frequencies, $\Omega = \sqrt{\omega_1 \omega_2}$ (increasing from left to right). Color bars are expressed in units of the power quantum $h f_1 f_2$. We consider commensurate drives with $r=2/3$.}
	\label{fig6}
\end{figure*}

In this Appendix we examine how the profile of frequency conversion and dissipation for electrons
evolves as a function of frequency. As explained in the main text, we increase the driving frequencies $\omega_j$ in tandem with the incidence angles of the modes, $\theta_j$, in such a way that $\cos \theta_j / \omega_j$ remains constant. This ensures that the surface ${\cal S}_{\balpha}$ in $\balpha$-space, traversed by electrons along their trajectories, remains fixed. In our analysis, we keep the ratio of frequencies $r=\omega_1/\omega_2$ constant, and vary their geometric mean $\Omega = \sqrt{\omega_1 \omega_2}$.

We contrast incommensurate ($r=1/\sqrt{2}$, Fig.~\ref{fig5}) and commensurate ($r=2/3$, Fig.~\ref{fig6}) driving. In both cases, the quantized conversion plateaus gradually shrink with $\Omega$, due to the expansion of Brillouin zone areas dominated by non-adiabatic effects including, in particular, Landau-Zener absorption. We find that the energy dissipation rate increases roughly linearly with frequency, and therefore becomes progressively less visible when normalized by the quantum of frequency conversion  $h f_1 f_2$. 

Periodic driving (Fig.~\ref{fig6}) possesses two main advantages over quasi-periodic driving (Fig.~\ref{fig5}), as discussed in the main text:  frequency conversion  persists at higher frequencies (panels a--e), and the total dissipated power is  smaller (panels f--j). As a result, commensurate driving supports much stronger net conversion at THz-scale frequencies---as demonstrated in Fig.~\ref{fig3} of the main text.

\section{Estimate of dissipation rate with a non-zero chemical potential}

Here we estimate the contribution to dissipation from a state with a non-zero chemical potential $\mu$---{that is, from the presence of an equilibrium Fermi surface}. We consider the regime where the characteristic relaxation time is much larger than the oscillation period of the driving modes, $\omega_j \tau \gg 1$ and neglect the effects of the perpendicular displacement field on the band dispersion.
We moreover assume $\mu$ to be small, working to leading order in $\mu$.
Without loss of generality, we consider the case where $\mu>0$ (negative $\mu$
is related to our results below through the approximate particle hole symmetry of the band structure of RMG).

 Eq.~\eqref{eq:master_equation} of the main text implies that the dissipation intensity is given by
\begin{equation}
 \eta_{\rm diss}(t) = \frac{\langle \mathcal E(t)\rangle - \langle  \mathcal{E}_{\rm eq}(t) \rangle}{\tau},
\end{equation}
with 
\newcommand{\idk}{\int\!\frac{d^2\kv}{4\pi^2}\, }
\begin{eqnarray}
    \langle \mathcal E(t)\rangle  &=& \idk  {\rm Tr}[\hat \rho(\kv,t)\hat H(\kv,t)], \\ 
    \langle   \mathcal{E}_{\rm eq}(t) \rangle &=& \idk {\rm Tr} [\hat \rho_{\rm eq}(\kv,t)\hat H(\kv,t)],
    \label{eqa:eeq_integral}
\end{eqnarray}
denoting the instantaneous and equilibrium energy densities, respectively. 
Here $\hat {\cdot}$ indicate operators acting on 
the Fock space of fermions with a given wavevector $\kv$. 

Since $ \hat H(\kv,t)=\hat H(\kv+e\vec A(t)/\hbar)$, we have  $\hat \rho_{\rm eq}(\kv,t)=\hat \rho_{\rm eq}(\kv+e\vec A(t)/\hbar)$. 
Thus, a shift of integration variable in Eq.~\eqref{eqa:eeq_integral},  $\kv\to \kv+e\vec A(t)/\hbar$ shows that  $\langle \mathcal{E}_{\rm eq}(t)\rangle$ is  time-independent and given by 
\begin{equation}
 \mathcal{E}_{\rm eq} = \sum_n \int \!\!\frac{d^2\kv'}{4\pi^2} 
f_{\rm \beta }(\varepsilon_n (\kv')- \mu)\varepsilon_n (\kv').
\end{equation}
where $f_{\beta}(\varepsilon)=1/(1+e^{\beta \varepsilon})$ denotes the Fermi-Dirac distribution and $\varepsilon_n(\kv)$ denote the energies of $H(\kv,t)$ for $n=0,1$. 
In the above we used  $\kv'$ for the integration variable, to make the distinction between ``stationary'' and ``comoving'' wavevectors clear. 

The goal of this Appendix is to compute the contribution to dissipation from intra-band scattering. 
Inter-band scattering in the form of Landau-Zener absorption is discussed in the main text and 
arises from nonadiabatic effects. We can hence isolate the dissipation from intra-band scattering by assuming adiabatic dynamics in our treatment~\cite{Nathan2022}.

Assuming adiabatic dynamics, and working in the limit $\omega_j \tau \gg 1$, $\rho(\kv,t)$ is diagonal in the eigenbasis of $H$, with stationary eigenvalues $ p_n(\kv)=\lim_{T\to \infty}\frac{1}{T}\int_0^T dt f_\beta(\varepsilon_n(\kv+e\vec A(t)/\hbar)$~\cite{Nathan2022}. 
As a result, 
\newcommand{\idt}{\lim_{T\to \infty} \int_0^T\,}
\begin{equation}
\langle \mathcal E(t)\rangle =  \sum_n  \idk  p_n(\kv) \varepsilon_n(\kv+e\vec A(t)/\hbar)
\end{equation}

The  average dissipation from intra-band relaxation   is given by 
\begin{equation}
    \bar \eta_{\rm mr} = \frac{1}{\tau}(\bar {\mathcal E} -\mathcal{E}_{\rm eq}),
\end{equation}
with $\bar{\mathcal E} \equiv \lim_{T\to \infty}\int_0^{T}\frac{dt}{T}\langle  \mathcal E(t)\rangle $ denoting the average energy density of the system.
We compute this quantity by  inserting the expression for $p_n(\kv)$ in the definition for $\bar {\mathcal E}$. 
Using  a change integration variables, $\kv \to \kv - e\vec A(t)/\hbar$, we obtain
\begin{equation}
 \bar {\mathcal E} = \sum_n  \idk \bar p_i(\kv) \varepsilon_n(\kv),
\end{equation}
where
\begin{widetext}
\begin{equation}
\bar p_n(\kv)= \lim_{T_i\to \infty}\int_0^{T_1}\frac{dt_1}{T_1} \int _0^{T_2}\frac{dt_2}{T_2} f_\beta\left(\varepsilon_n\left(\kv+\frac{e}{\hbar}[\vec A(t_1)-\vec A(t_2)]\right)-\mu\right).
\end{equation}
\end{widetext}
Note that $p_i(\kv)$ is obtained from effectively averaging the equilibrium distribution $f_\beta(\varepsilon_i(\kv)-\mu)$ over $\kv$-points reached by the $\kv$-displacements in $\{\frac{e}{\hbar}[\vec A(t_1)-\vec A(t_2)]|t_1,t_2\in[0,\infty)\}$.
Using this result, we obtain
\begin{equation}
 \bar \eta_{\rm mr} =\frac{1}{\tau} \sum_n  \idk \varepsilon_n(\kv)[ \bar p_n(\kv) -f_{\beta }(\varepsilon_n-\mu)].
\end{equation}
 
To estimate $\bar \eta _{\rm mr}$, we approximate $\bar p_0(\kv)=1$, and assume the  valence band carriers  to be uniformly smeared over a disk of radius $ \Delta k = 3 e A_{\parallel}/2\hbar$ with $A_{\parallel}$ denoting the characteristic scale of the  in-plane vector potential from  the drive (recall that we consider the limit of small $\mu$, where $k_{\rm F}$ in equilibrium is much smaller than $\Delta k$).
This results  in  $\bar p_1(\kv) = \frac{k_{\rm F}^2}{\Delta k^2}\theta(\Delta k-|\kv|)$  with $\theta$ denoting the Heaviside step function.
The  estimate of $\bar p_1(\kv)$  is based on the fact that $\bar p(\kv)$ has maximum at $|\kv|=0$, decreases monotonically with  $|\kv|$, and vanishes for  $|\kv|\geq 2eA_{\parallel}/\hbar$.
We finally assume $f_{\beta}(\varepsilon- \mu )\propto \theta( - (\varepsilon - \mu))$.

\begin{figure*}[t]
	\includegraphics[width=0.98\textwidth]{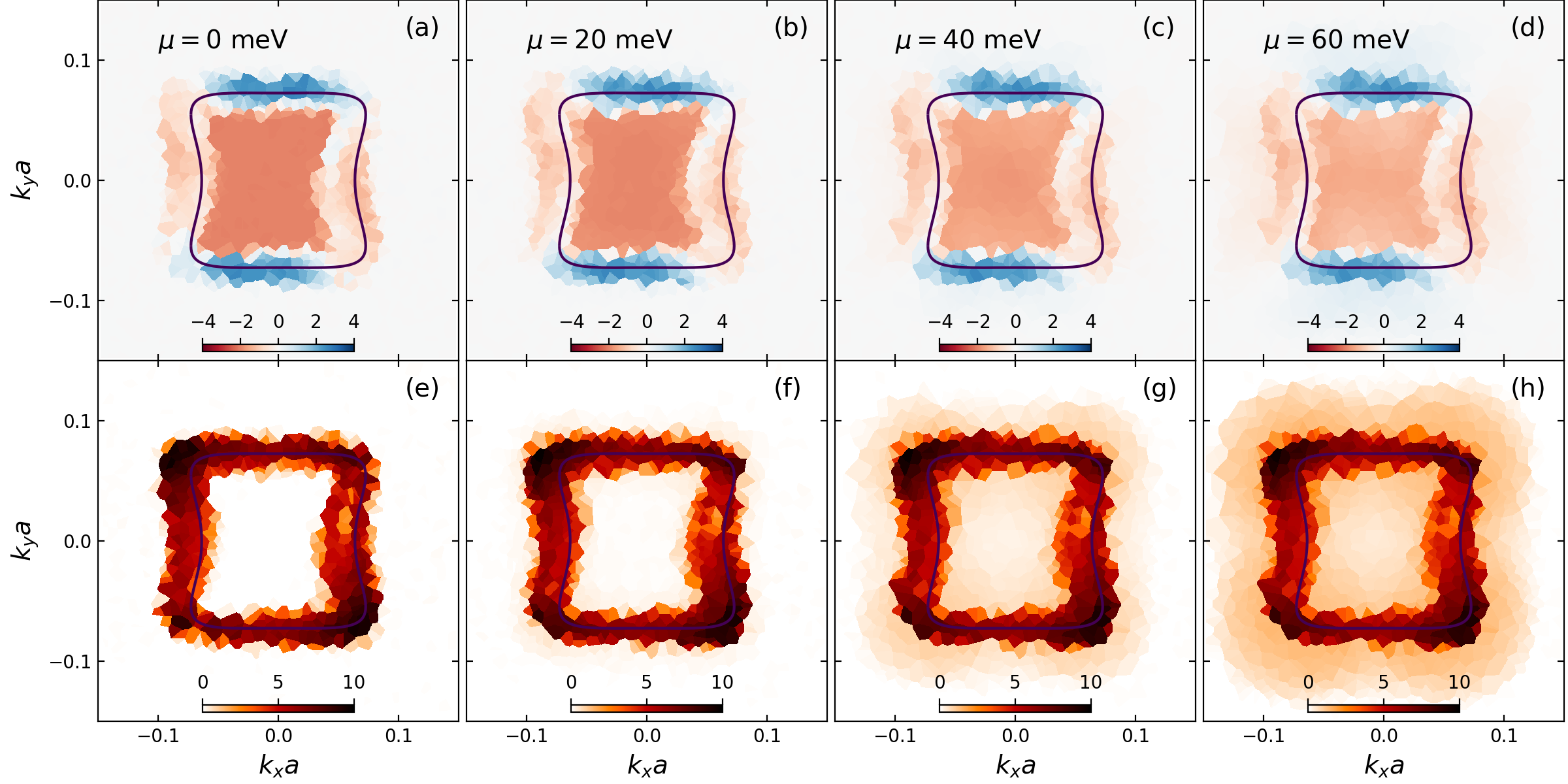}
	\caption{Scaling of the (time-averaged) frequency conversion power  $\cP^+_{\rm FC}$ (panels a--d), and the corresponding dissipated power  $\cP^+_{\rm diss}$ (panels e--h), as a function of the chemical potential $\mu$. Color bars are expressed in units of the power quantum $h f_1 f_2$. We consider incommensurate frequencies with ratio $r=1/\sqrt{2}$ and parameters identical to those used for Fig.~\ref{fig2}a--c in the main text. As the chemical potential is increased, two trends are observed. First, the quantization of the frequency conversion power is gradually lost due to the opposite contribution of the conduction band (panels a--d). Second, the dissipation profile acquires a diffuse contribution from the underlying Fermi surface (panels e--h), which coexists with the sharp ``ring of fire'' due to Landau-Zener absorption.}
	\label{fig7}
\end{figure*}

Using the above assumptions along with  $\varepsilon_1(\kv)={\gamma_L} \kv^L$ and $\varepsilon_0(\kv)=-{\gamma_L} \kv^L$, we obtain
\begin{equation}
\bar \eta_{\rm mr}\approx \idk \frac{\gamma_{L}}{\tau} \kv^L \left[\theta(\Delta-|\kv|)\frac{k_{\rm F}^2}{\Delta k^2}-\theta(k_{\rm F}-|\kv|)\right].
\end{equation}
Going to polar coordinates, $\kv = k(\cos\phi,\sin\phi)$, we obtain 
\begin{equation}
\bar \eta_{\rm mr}\approx  \frac{ {\gamma}_{L}}{2\pi\tau}\int_{0}^{\infty} \!\!\! dk\,    k^{L+1} \left[\theta(\Delta k-k)\frac{k_{\rm F}^2}{\Delta k^2} -\theta(k_{\rm F}-k)\right].
\end{equation}
Evaluating the integral and using $\mu = \gamma_L k_{\rm F}^L$ we find 
\begin{equation}
\bar \eta_{\rm mr}\approx  \frac{\mu k_{\rm F}^2}{2\pi\tau(L+2)} [(\Delta k/k_{\rm F})^{L} -1].
\end{equation}
Using our estimate $\Delta k\approx 3eA_{\parallel}/2\hbar$, we thus obtain 
\begin{equation}
\bar \eta_{\rm mr}\approx  \frac{(\mu/\gamma_L)^{2/L}}{2\pi\tau(L+2)} \left[\gamma_L \left(\frac{3eA_{\parallel}}{2\hbar}\right)^L - \mu \right].
\end{equation}
Working to leading order in $\mu$ yields
\begin{equation}
\bar \eta_{\rm mr}\approx \mu^{2/L} \frac{ \gamma_L^{2/L-1}}{2\pi\tau(L+2)} \left(\frac{3eA_{\parallel}}{\hbar}\right)^L .
\end{equation}
We next use $A_\parallel  \sim 2A \cos\theta \sim 2E \cos \theta/\omega$ with  $A$ and $E$  denoting the characteristic amplitude of the {\it total} vector potential and electric field from  each drive, respectively, leading to
\begin{equation}
\bar \eta_{\rm mr}\approx  \frac{ \mu^{2/L}\gamma_L^{1-2/L}}{2\pi\tau(L+2)} \left(\frac{3 eE\cos\theta}{\hbar\omega}\right)^Lx .
\end{equation}
For $L=2$, this expression is particularly simple: 
\begin{equation} 
\bar \eta_{\rm mr}\approx  \frac{\mu}{8\pi\tau} \left(\frac{3eE\cos\theta}{\hbar\omega}\right)^2 \quad (L=2).
\label{eqa:mr_estimate}
\end{equation}

\begin{figure*}[t]
	\includegraphics[width=0.6\textwidth]{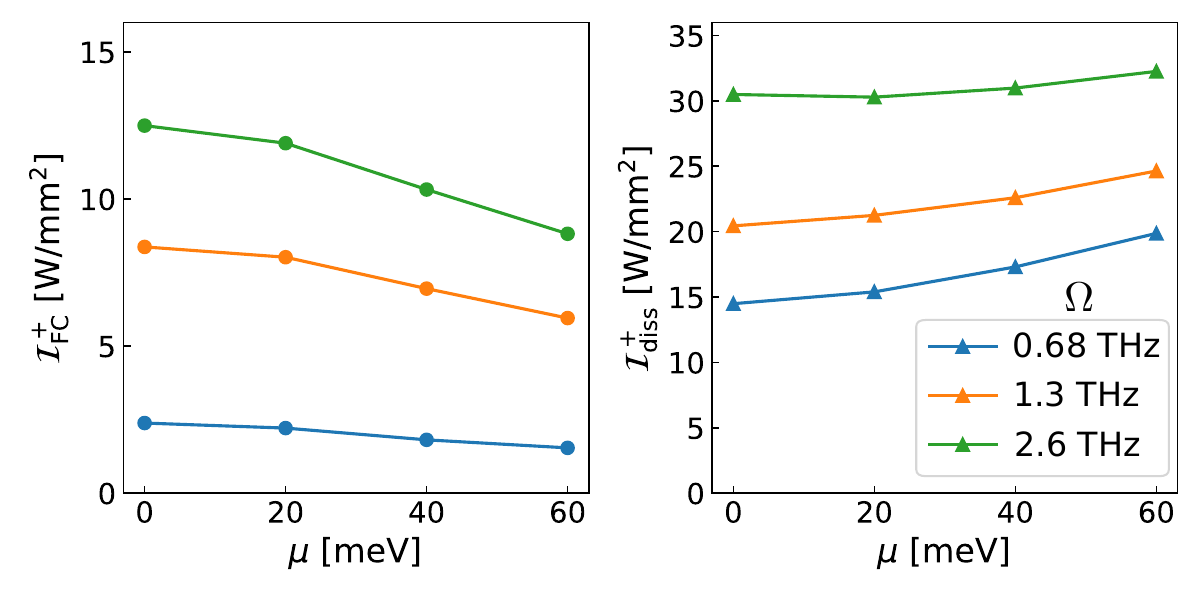}
	\caption{Scaling of the valley-resolved frequency conversion intensity $\cI^+_{\rm FC}$ (left) and the corresponding dissipation intensity $\cI^+_{\rm diss}$ (right), as a function of the chemical potential $\mu$ and different frequencies $\Omega = \sqrt{\omega_1 \omega_2}$, with an incommensurate ratio $r=1/\sqrt{2}$ as in Figs.~\ref{fig2}, \ref{fig5} and \ref{fig7}. The data for the lowest $\Omega$ are obtained by integrating the power distributions shown in Fig.~\ref{fig7} (a-d) and (e-h), respectively.}
	\label{fig8}
\end{figure*}

We now estimate the dissipation caused by the Fermi surface from this expression, and compare with numerical simulations carried out at non-zero $\mu$ in Figs.~\ref{fig7}, \ref{fig8}.
Note that Fig.~\ref{fig8} shows the {\it total} dissipated power, including both Landau-Zener absorption and momentum relaxation due to the presence of a Fermi surface. We expect that dissipation strength at $\mu=0$ serves as a good proxy for the characteristic rate of Landau-Zener absorption, while the difference in dissipation to the $\mu=0$ value, $\cI_{\rm diss}^+(\mu)-\cI_{\rm diss}^+(0)$ serves as a reasonable estimate for the momentum-relaxation induced dissipation, $\eta_{\rm mr}$.

Using frequencies $\omega \sim 0.68$ THz, electric fields $E \sim 0.1$ V/nm, relaxation time $\tau = 500$ ps, incidence angles $\theta_j - \pi/2 \approx 0.1^\circ$, our estimate in Eq.~\eqref{eqa:mr_estimate} yields $\eta_{\rm mr} \approx 0.5$ W/mm$^2$ for  $\mu \approx 40$ meV. This rough estimate agrees reasonably well with the numerical simulations shown in Fig.~\ref{fig8}, which suggest $\eta_{\rm mr}\sim 2$ W/mm$^2$.

This analysis also predicts that Fermi-surface-induced dissipation stays constant when scaling up the frequencies in conjunction with the incidence angle such that $\cos \theta_j/\omega_j$ remains fixed, as in Fig.~\ref{fig3} of the main text. We also verify this scaling in Fig.~\ref{fig8}. While the $\mu=0$ dissipation increases with frequency due to more prominent Landau-Zener tunneling, the  Fermi surface contribution (which turns on with $\mu$) is on the same scale and roughly independent of frequency. Note that we consider incommensurate frequencies; for commensurate frequencies, we expect the dissipation from Landau-Zener absorption to be further suppressed as demonstrated in Fig.~\ref{fig3}.

\section{Estimate of the heating rate from driving}

We finally estimate the heating rate from the driving configuration we consider in the main text, which will limit the  lifetime of  topological frequency conversion. From the numerical simulations in Fig.~\ref{fig3} we extract a typical scale for the dissipation intensity $I_{\rm diss} \sim 40$ W/mm$^2$ for commensurate driving with frequencies $f_i \sim1$ THz and $\tau = 500$ ps, accounting for dissipation from {\it both} valleys. The heat capacity of graphene in the temperature range $T \sim 10 - 100$K is dominated by phonons and of order $C \sim \left( 0.1 - 1 \right) \frac{\rm J}{\rm K\, mol}$~\cite{Pop2012}. Using the density of carbon atoms of $2/A_{\rm uc}$ with the unit cell area $A_{\rm u.c.} = 5.2\, {\rm \AA}^2$ yields the heat capacity per unit area $c \sim 10^{-11} - 10^{-12} \frac{\rm J}{\rm K\, mm^2}$. We use this value to estimate the heat capacity for bilayer graphene as $2c$, resulting in a heating rate given by  $\eta = \frac{I_{\rm diss}}{2 c} \sim 10^{11} - 10^{12}$ K/s. 
The phonon temperature in irradiated bilayer graphene is thus expected to increase by a Kelvin for every $1-10$ ps.  
This estimate should allow for multiple THz-frequency cycles before the temperature increase of phonons causes the  electronic lifetime $\tau$ to decrease substantially---thus resulting in the gradual disappearance of frequency conversion.

\end{document}